\documentclass[twocolumn,showpacs,preprintnumbers,aps,prb]{revtex4}

\usepackage{graphicx}
\usepackage{dcolumn}
\usepackage{bm}
\usepackage{subfigure}

\begin{document}
\title{Dielectric properties and spin-phonon coupling in multiferroic double perovskite
Bi$_2$CoMnO$_6$ }

\author{Satadeep Bhattacharjee, Olle Eriksson and Biplab Sanyal\cite{bs}}
\affiliation{Department of Physics and Astronomy, Box 516, 75120, Uppsala University, Uppsala, Sweden}
\date{\today}

\begin{abstract}
\noindent  First-principles electronic structure calculations have been performed for the double perovskite Bi$_2$CoMnO$_6$  in its non-centrosymmetric polar state using generalized gradient approximation plus Hubbard U approach. We find that while Co is in a high spin state, Mn is
in a intermediate spin state. The calculated dynamical charge tensors are anisotropic reflecting a low-symmetry structure of the compound. Magnetic structure dependent phonon frequencies indicate the presence of spin-phonon coupling. Using Berry phase method, we obtain a spontaneous electronic polarization of 5.88 ${\mu}C/cm^2$ which is close to the experimental value 
observed for a similar compound, Bi$_2$NiMnO$_6$.
\end{abstract}

\vspace{20mm}
\pacs{75.50.Pp,75.70.-i,71.70.Gm}

\maketitle
Magnetic double perovskites with formula A$_2$BB$^\prime$O$_6$ (A is an alkaline rare-earth, B and B$^\prime$ are transition metals) have attracted a lot of attention in the last decade due to their potential in spintronic applications specially since the discovery of low field magneto-resistance in Sr$_2$FeMoO$_6$ and Sr$_2$FeReO$_6$.\cite{dp01}. Depending on the stoichiometry and temperature, a number of double perovskites show inter-related magnetic, structural and electronic phases. Although, many of these materials order ferromagnetically at high temperature, very few of them show ferroelectricity \cite{dp1,dp2,dp3,dp4,dp5}. A perpetual interest exists to identify materials having multiferroic properties with the simultaneous existence of ferroelectric and magnetic order \cite{mfrev}.
Here we have studied Bi$_2$CoMnO$_6$ that crystallizes in the double perovskite structure. It is among a few, rare materials which show existence of both ferroelectric (FE) as well as ferromagnetic (FM) behavior. Due to the presence of 6s$^2$ lone pair of Bi$^{3+}$, there is a non-centrosymmetric (space group C2 (no. 5)) structural distortion which leads to a FE transition temperature around 600 K\cite{JJSPM}, above which the centrosymmetric structure appears with space group C$2/m$(no. 12). Below the magnetic ordering temperature of 95 K, ferromagnetism appears due to the 180$^{\circ}$ 
cation-anion-cation ferromagnetic superexchange interaction\cite{th_P1} in the chain Co$^{+}$-O$^{-}$-Mn$^{+}$ where Co and Mn ions are ordered in a rock-salt configuration. A recent experimental study has claimed a strong dependence of T$_c$ on the monoclinic angle\cite{Mang1}. 
\begin{figure}[h]
\includegraphics{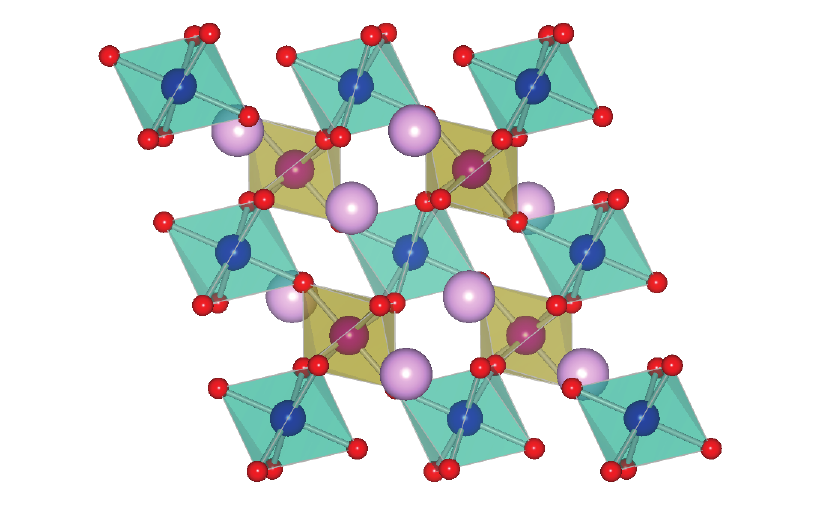}
\caption{(Color online) Crystal structure of BCMO, as viewed along the short-axis {\bf b}, showing octahedral tilt patterns of CoO$_6$ (blue) and MnO$_6$ (red) octahedra along with Bi as purple balls. }
\label{fig1}
\end{figure}
In the present study, we concentrate on the C2 structure of Bi$_2$CoMnO$_6$ (BCMO) as first-principles calculations have not been reported in the literature for either
of the two structural phases. 
Calculations were performed using density functional
theory (DFT) within the Perdew-Burke-Ernzerhof generalized gradient approximation (GGA) on the basis
of the projector augmented wave method implemented in the Vienna \textit{ab initio}
simulation package (VASP)\cite{vasp}. The wave function was expanded in plane waves using 500 eV energy cut-off. For Bi and Mn, we have used pseudopotentials with semi-core electrons. Brillouin zone integrations were carried out using a $4\times6\times4$ Monkhorst-Pack $k$-point mesh. To treat strong electron interaction effect in Co and Mn we have carried out GGA+U calculation in the Hubbard model framework with the choice of Coulomb parameter U=3 eV for Co-d and U=4 eV for Mn-d orbitals. The value of the exchange parameter J used was 1 eV for both the transition metals. The unit cell consists of four formula units in the C$2$ phase. In Wyckoff notations, the four Co atoms occupy 2a and 2b sites respectively while Bi, Mn and O occupy 4c sites. In Glazer's notation, the tilt pattern of the octahedra is given by a$^-$b$^0$c$^-$ (Fig.~1) which imply that there is no tilting of octahedra along [010] direction, while the neighboring octahedra tilt in opposite directions to each other along both [100] and [001] directions. We have performed the structural optimization involving the relaxation of ionic positions as well as cell shape and volume. Forces were relaxed up to values smaller than 0.005 eV/\AA. For the calculation of density of states we used a  $k$-point mesh of $6\times6\times6$ with 54 k-points in the irreducible Brillouin zone. The dielectric and dynamical properties such as Born
effective charge (BEC) matrices and zon-center phonon frequencies were calculated using density functional perturbation theory (DFPT)\cite{dfpt}.

It should be noted that we have used GGA+U method in the present study as GGA calculations are unable to yield an insulating state. Our optimized lattice parameters are a= 9.62 \AA, b= 5.50 \AA~and c= 9.74 \AA (unit cell volume of 490.21 \AA$^3$). The corresponding experimental values are 9.49 \AA, 5.45 \AA~and 9.57 {\AA} respectively\cite{JJSPM} (unit cell volume of 471.13 \AA$^3$). The calculated monoclinic angle $\beta$ between vectors {\bf a} and {\bf c}, is $108.15^{\circ}$, which is slightly overestimated with respect to the experimental one ($107.60^{\circ}$). 
\begin{figure}[h]
\includegraphics[scale=0.35]{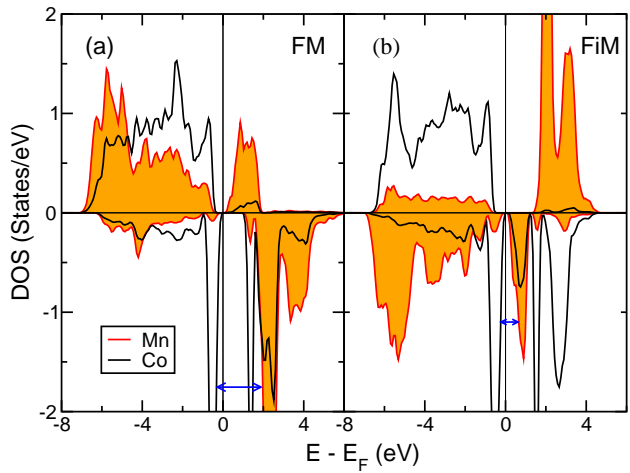}
 \caption{(Color online) Mn and Co DOSs projected on d-orbitals for (a) FM and (b) FiM configurations. The horizontal blue lines indicate separation between the occupied Co and unoccupied Mn states in the minority spin. See text for details.}
\label{fig2}
\end{figure}

In Fig.~2, we show the density of states (DOS)
obtained by GGA+U method for the ferromagnetic and ferrimagnetic (FiM) states. The choice of our U values opens up a band gap of around 0.3 eV in the FM state. The top of the valence band is dominated by Co-d and O-p states (not shown) while the Mn states are pushed about 1 eV below the Fermi energy. On the other hand the bottom of the conduction band is due to Mn-d and O-2p states. 
A comparison of the total energies of iso-structural FM and FiM structures yields that the FM state is 163 meV/f.u. lower in energy compared to the FiM state. The magnetic moment of Co (both in 2a and 2b sites) is found to be 2.5 $\mu_B$, while Mn carries a magnetic moment of 3.2 $\mu_B$, yielding a magnetic moment of 6 $\mu_B$/f.u. in the FM state. The analysis of orbital projected charges in both FM and FiM magnetic structures shows that Co is in a high spin state with around 7 $d$ electrons (Co$^{2+}$ state), while the Mn sphere contains around 5 $d$ electrons. It seems that Mn is in an oxidation state which is associated with an ionicity of about 2+ rather than 4+, which is the expected ionic state from the counting of oxidation states. This deviation from the expected ionicity may be attributed to the low symmetry structure and small Mn-O bond lengths. Similar ill-defined oxidation states of Mn and O were mentioned by Ciucivara {\it et. al} \cite{dp4} for an iso-structural compound Bi$_2$NiMnO$_6$. The structural analysis shows that in the FM state, ${\angle{Mn-O-Co(2a)}}$ is  equal to 159.45$^{\circ}$ while the angle ${\angle{Mn-O-Co(2b)}}$ is 150.83$^{\circ}$ giving an average ${\angle{Mn-O-Co}}$ angle of 155.14$^{\circ}$. In both the Mn-O-Co chains, the Co-O bond is about 0.18 {\AA}  longer than the Mn-O bond. This clearly indicates that the Mn-O bond is stronger in comparison with Co-O bond, which is reflected in the DOS where we see that more Mn-d states are pushed down below the  Fermi level.

We have computed a spontaneous electronic polarization of 5.88 ${\mu}C/cm^2$ in the FM state by Berry phase method. A $4\times4$ array of strings containing 32 k-points has been used to obtain the electronic polarization, which is non-zero only along the short-axis {\bf b} in agreement with the C2 symmetry of the crystal. Our estimated value is close to what has been measured for Bi$_2$NiMnO$_6$ thin films. Unfortunately, we are unable to find an experimental value for BCMO.

\begin{table}
\begin{center}
\begin{tabular}{cccc}
$Z^{*}_{Bi}: $ & $\left( \begin{tabular}{ccc}
4.51 &  0.18 & 0.32 \\
0.21 &  4.71 &   -0.35\\
-0.14  &  -0.10 &    4.98 \\
\end{tabular}
\right)$
,
$\left( \begin{tabular}{ccc}
4.55 &    0.18 &    0.30 \\
0.26 &    4.77 &   -0.41 \\
-0.09 &   0.0 &    4.95 \\
\end{tabular}
\right)$
\\
$Z^{*}_{Co}: $ & $\left( \begin{tabular}{ccc}
2.81 &    0.00 &     0.26 \\
0.00 &     2.47 &     0.00 \\
\fbox{0.02} &   0.00 &    2.84 \\
\end{tabular}
\right)$
,
$\left( \begin{tabular}{ccc}
2.83 &    0.00 &     0.34 \\
0.00 &     1.64 &     0.00 \\
\fbox{1.15}&    0.00 &     2.74 \\
\end{tabular}
\right)$
\\
$Z^{*}_{Mn}: $ & $\left( \begin{tabular}{ccc}
3.31 &     0.58 &    -0.32 \\
0.74 &     4.25 &     0.47 \\
-0.37 &     \fbox{0.07}&     4.05 \\
\end{tabular}
\right)$
,
$\left( \begin{tabular}{ccc}
3.80 &     0.29 &    -0.41 \\
0.67 &     4.05 &     0.17 \\
0.21 &     \fbox{0.62}&     4.39 \\
\end{tabular}
\right)$
\\
$Z^{*}_{O}: $ & $\left( \begin{tabular}{ccc}
-3.37 &    -0.19 &    -0.14  \\
-0.20 &    -2.28 &     0.14 \\
0.33 &     0.57 &    -2.26 \\
\end{tabular}
\right)$
,
$\left( \begin{tabular}{ccc}
-3.35 &     0.08 &     0.06  \\
-0.18 &    -2.27 &     0.15  \\
 0.23 &     0.42 &    -2.29 \\
\end{tabular}
\right)$
\\
\end{tabular}
\end{center}
\caption{Born effective charges of Bi, Co, Mn and O (one in the Co-O-Mn chain)  for FM (left) and FiM (right) configurations.}
\end{table}
\par
Now we discuss the effect of magnetic structure on the BEC matrices shown in Table~1. It is observed that they are quite anisotropic due to the highly distorted structure of the compound and there are significant differences in BEC's for FM and FiM states for Co and Mn. Both diagonal and off-diagonal components of BEC matrices of Co and Mn are different in FM and FiM states, with major changes observed for ($Z^*)_{31}$ and ($Z^*)_{32}$ components (shown in box) of Co and Mn BEC's respectively. These changes may be attributed to the changes in electronic DOS's in two magnetic states, where, e.g., the energy separation between the occupied Co and unoccupied Mn $t_{2g}$ states in the minority spin channel (Fig.~2) is decreased significantly in FiM state compared to the FM state. According to Ghosez {\it et al.} \cite{ghosez}, the expression for BEC has the energy separation in the denominator and hence a reduction of this value should increase BEC significantly. A relatively small change occurs in the diagonal components of Mn as the Mn-O bonding is almost uniform over a wide range of energy whereas a significant Co-O hybridization about 0.5 below the Fermi energy increases the diagonal component of Co BEC's. The BEC's of O are almost identical in both structures. Among all the species, Bi has the largest BEC, demonstrating a strong Bi-O covalancy.  
From all these observations we can conclude a significant magnetism dependent electrostructural coupling in the system very similar to what has been observed by Das et. al \cite{Hena} for La$_2$NiMnO$_6$. However unlike the study in Ref.~\onlinecite{Hena}, we did not observe an anomalously large off-diagonal BEC for either of the magnetic specie. 
\begin{figure}[h]
\includegraphics[scale=0.35]{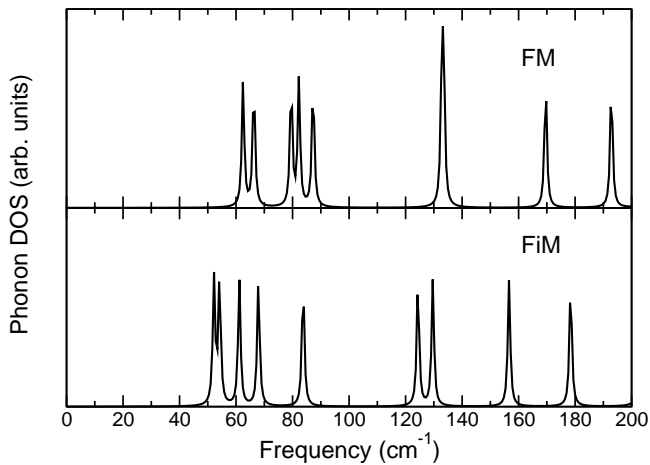}
 \caption{Phonon DOS of IR-active modes for FM and FiM configurations.}
\label{fig3}
\end{figure}

Finally, we analyze the zone center phonon modes for FM and FiM structures. The eight lowest infra-red (IR) active modes are shown in Fig.~3. One can observe that the typical softening of IR-active modes could be as large as 10 cm$^{-1}$. The frequencies at 62, 66, 79, 82, 87, 133,133 and 169 cm$^{-1}$ in the FM phase are shifted to 52, 54, 61, 67, 83, 124, 129 and 156  cm$^{-1}$ in the FiM phase. The least shifted mode is  at 133 cm$^{-1}$ where the softening is associated with the further lifting of degeneracy of this mode. Subsequently there is an increase in the static dielectric constant in FiM phase. To understand the spin-phonon coupling better, we analyzed the softest mode (at 52 cm$^{-1}$) in the FiM state more carefully. By adding the atomic displacement due to this mode to the equilibrium atomic positions we find that the atomic displacements  are such that the increase in Mn-O-Co angle with respect to FM state is very small. We find an average increase of about 2$^{\circ}$ of Mn-O-Co angle. This fact may contribute to the observation that the spin-phonon coupling though present, is small. Moreover, the contribution of Mn to the change in mode effective charges of the IR-active modes of two magnetic states as well as to dielectric constant is smaller compared to Co.

In summary, our GGA+U calculations reveal that Co and Mn are in high and intermediate spin states respectively in the double perovskite Bi$_2$CoMnO$_6$ with a calculated band gap of 0.3 eV. The computed dynamical charge matrices are highly anisotropic due to the low symmetry structure. Magnetic structure dependent phonon frequencies indicate a spin-phonon coupling in this compound. Our theoretically estimated spontaneous polarization of 5.88 ${\mu}C/cm^2$ is close to the experimental value observed for the thin film of a similar compound, Bi$_2$NiMnO$_6$.

We acknowledge Swedish Research Council for financial support and Swedish National Infrastructure for Computing (SNIC) for the allocation of supercomputing time. O.E. acknowledges support from ERC. We thank Nicola Spaldin for valuable discussions. 
  

\end{document}